# Lattice Instability and Competing Spin Structures in the Double Perovskite Insulator Sr$_2$FeOsO$_6$

Avijit Kumar Paul[a,b], Manfred Reehuis[c], Vadim Ksenofontov[d], Binghai Yan[a], Andreas Hoser[c], Daniel M. Többens[c], Peter Adler[a,*], Martin Jansen[a,b,†], and Claudia Felser[a,d]

[a]*Max-Planck-Institut für Chemische Physik fester Stoffe, 01187 Dresden, Germany,*

[b]*Max-Planck-Institut für Festkörperforschung, 70569 Stuttgart, Germany,*

[c]*Helmholtz-Zentrum für Materialien und Energie, 14109 Berlin, Germany*

[d]*Institut für Anorganische Chemie und Analytische Chemie, Johannes Gutenberg-Universität, 55128 Mainz, Germany*

The semiconductor Sr$_2$FeOsO$_6$, depending on temperature, adopts two types of spin structures that differ in the spin sequence of ferrimagnetic iron - osmium layers along the tetragonal *c*-axis. Neutron powder diffraction experiments, $^{57}$Fe Mössbauer spectra, and density-functional theory calculations suggest that this behavior arises because a lattice instability resulting in alternating iron-osmium distances fine-tunes the balance of competing exchange interactions. Thus, Sr$_2$FeOsO$_6$ is an example for a double perovskite, in which the electronic phases are controlled by the interplay of spin, orbital, and lattice degrees of freedom.

PACS numbers: 75.25.-j, 75.50.Ee, 76.80.+y

Transition metal (TM) double perovskites (DPs) $A_2BB'O_6$ in which $A$ is an alkaline, alkaline earth, or rare earth atom feature ordered arrangements of two different corner-sharing TM $BO_6$ and $B'O_6$ octahedra. Compared to that of simple $ABO_3$ perovskites, this type of atomic order offers ample possibilities for the design of magnetic materials [1,2], especially if 3$d$ TM ions at $B$ with essentially localized strongly correlated electrons are combined with more electron-itinerant 4$d$ or 5$d$ ions at $B'$. Illustrative examples are the half-metallic DPs $Sr_2FeMoO_6$ [3] and $Sr_2FeReO_6$ [4], which have remarkably high ferromagnetic transition temperatures $T_C$ around 400 K and large tunnel magnetoresistance (TMR) effects even at room temperature and in low magnetic fields. As these materials are potential candidates for spintronic applications [5], the properties of $Sr_2FeMoO_6$ and $Sr_2FeReO_6$ have been studied extensively [6]. Their metallic behavior and ferromagnetism are explained by an interaction between $B$ and $B'$ sites that can be considered as a generalized double exchange [7] involving a hybridization-driven (HD) delocalization of $t_{2g}$ minority spin electrons in a hybridized $t_{2g}(B)\downarrow$-O-$t_{2g}(B')\downarrow$ band [8-10]. The majority spin electrons are essentially localized at the Fe sites and give rise to a $t_{2g}^3\uparrow e_g^2\uparrow$ configuration. Since there is an excitation gap in the majority spin channel, these materials are half-metallic.

Most remarkably, in the $Sr_2CrB'O_6$ series of Cr-based DPs, even higher $T_C$ values have been found ranging from 458 K for $B'$ = W [11], to 635 K for $B'$ = Re [12], to 725 K for $B'$ = Os [13]. In this series, the number of valence electrons introduced by the 5$d$ element increases from 1 for W, to 2 for Re, to 3 for Os. While $Sr_2CrWO_6$ and $Sr_2CrReO_6$ are metallic, $Sr_2CrOsO_6$, which has the highest $T_C$, is a ferrimagnetic insulator. This is difficult to explain with the HD mechanism, which should be less effective in $Sr_2CrOsO_6$ because of the decreased $B$-$B'$ hybridization [14]. Recent band structure and model Hamiltonian calculations suggest that instead, an interplay between the HD mechanism and superexchange (SE) interactions between local Cr and Os magnetic moments leads to stabilization of a ferrimagnetic [14] or canted [15] magnetic state with high $T_C$.

In this Letter, we consider the iron-based DP series $Sr_2FeB'O_6$. Whereas $Sr_2FeWO_6$ (SFWO) is known to be an antiferromagnetic insulator with $T_N \sim 30$ K with $Fe^{2+}$ and $W^{6+}$ sites [16], $Sr_2FeReO_6$ is a metallic mixed-valence ferromagnet [4,17]. The Os-containing $Sr_2FeOsO_6$ has previously been discussed only briefly [18,15]. In a theoretical study [19], it was considered to be a *ferrimagnetic* semiconductor like $Sr_2CrOsO_6$. However, recently, we [20] and another group [21] showed that $Sr_2FeOsO_6$ is an *antiferromagnetic* semiconductor. Magnetic susceptibility and heat capacity measurements indicated two magnetic phase transitions at 140 K and 67 K [20]. Here, we report a comprehensive investigation of the spin structures, the magnetic phase transitions, magnetostructural correlations, and the electronic structure in $Sr_2FeOsO_6$ using temperature-dependent neutron powder (NP) diffraction experiments, $^{57}$Fe Mössbauer spectroscopy, and density functional theory (DFT) calculations. In contrast to $Sr_2CrOsO_6$, $Sr_2FeOsO_6$ is found to have two types of antiferromagnetic spin structures that differ only in the spin ordering pattern along the $c$-direction of the tetragonal crystal structure. Structural anomalies and the DFT results point toward a lattice instability involving alternating Fe-Os distances, which fine-tunes the competing exchange interactions. Accordingly, $Sr_2FeOsO_6$ is an example material demonstrating the intimate interplay of lattice, spin, and orbital degrees of freedom, and possibly electron correlation effects in a double perovskite with a well ordered atomic arrangement of rather localized $3d$ and more delocalized $5d$ sites.

All NP patterns collected between 2 and 300 K could be successfully refined in the tetragonal space group $I4/m$ (No. 87) [22], which is consistent with our previous powder X-ray results [20]. The bond distances at 2 K reveal that both the $FeO_6$ and the $OsO_6$ octahedra are almost regular. The osmium – oxygen bond distances are very close to the value of 1.957 Å found earlier for the trigonal double perovskite $Sr_2CrOsO_6$ (space group $R\bar{3}$) at 2 K [13]. For both compounds, it can be concluded that the valence states are +5 for Os and +3 for Cr or Fe. From 300 to 2 K, the lattice parameters $a$ and $c$ become considerably shorter and longer, respectively, as shown in Fig. 1. In cubic notation, the $c/a$ ratio increases from 1.00558(7) at 300 K to 1.01689(7) at 2 K. The increase in the tetragonal distortion is strongest between 140 and 67 K, where an overall antiferromagnetic spin structure evolves.

Below the Néel temperature $T_N$ = 140 K, magnetic intensity is observed at the position of the reflection $(1,0,0)_M$, which is forbidden for the space group $I4/m$. Thus, the magnetic moments of the metal atoms connected via the translation $t = (½,½,½)$ are coupled antiparallel to each other. This magnetic phase is denoted as AF1. No magnetic intensity was observed at the position of the reflection $(0,0,1)$, and hence, the moments must be aligned parallel to the tetragonal $c$-axis (Fig. 1e, left). At 75 K, the ordered moment of the iron atoms $\mu_{exp}(Fe)$ = 1.83(7) $\mu_B$ was found to be much larger than that of the osmium atom $\mu_{exp}(Os)$ = 0.48(9) $\mu_B$. Below $T_2$ = 67(2) K, the magnetic intensity of the $(1,0,0)_M$ reflection spontaneously decreased (Fig. 1b). In addition, a second set of magnetic reflections appeared, which can be generated with the propagation vector $k_2$ = $(0,0,½)$. The absence of the reflection $(0,0,½)_M$ indicates that the magnetic moments are again aligned parallel to the $c$-axis (Fig. 1e, right), and at 2 K as well, the ordered moment of the iron atoms $\mu_{exp}(Fe)$ = 3.07(5) $\mu_B$ is larger than that of the osmium atom $\mu_{exp}(Os)$ = 0.68(9) $\mu_B$. Even at 2 K, a residual fraction of AF1 phase persists. Taking into account both magnetic phases, we calculated a total moment of $\mu_{tot}(Fe)$ = 3.34(7) $\mu_B$ and $\mu_{tot}(Os)$ = 0.71(11) $\mu_B$.

Since we used the overall scale factor of the crystal structure refinement for the calculation of the magnetic moments of each magnetic phase, we have not been able to obtain the individual magnetic moments of each phase in the coexisting state. Based on the Mössbauer results discussed later, we assume that Fe and Os have the same moment values in the two magnetic phases and find that the sample at 2 K contains 85(2) % of AF2 and 15(2) % of AF1. In both magnetic structures, the Fe and Os moments are coupled antiparallel to each other within the tetragonal $ab$-plane. In the AF1 phase, however, the Fe and Os atoms form ferromagnetic chains along the $c$-axis. This + + + + sequence along $c$ changes to the spin sequence + +, − −, + +, … of the Fe and Os atoms in the low-temperature AF2 phase. This possibly indicates the presence of a dimerization, where the Fe-Os separations alternate along the *c*-axis. Such dimerization would require a lower-symmetry space group than $I4/m$. Since our synchrotron and neutron diffraction studies did not show additional primitive Bragg reflections violating the *I*-centering, we refined the low-temperature crystal structure in space group $I4$ (No. 79) [22]. From this analysis, we finally obtained separations along the *z*-direction of $d$(Fe-Os) = 3.918(18) Å and $d$(Fe-Os) = 4.014(18) Å from the neutron data at 2 K or of $d$(Fe-Os) = 3.892(4) Å and $d$(Fe-Os) = 4.047(4) Å from the synchrotron data [20] at 15 K. Finally, it is noted that the magnetic moments in the AF1 phase evolve only gradually below $T_N$, while full order is reached within a narrow temperature range below $T_2$ (Fig. 1d).

The $^{57}$Fe Mössbauer spectra of $Sr_2FeOsO_6$ (Fig. 2a) give clear independent evidence for the occurrence of two magnetic phase transitions. Spectra measured at 295 and 150 K contain a single line without any apparent quadrupole splitting, whereas the spectra at lower temperatures show hyperfine sextets, which are typical for magnetically ordered Fe sites, and a small but finite quadrupole interaction. The latter is correlated with the magnetoelastic coupling reflected in the lattice parameters. Anomalies in the magnetic susceptibilities near 250 K (Fig. 2d) are not due to a magnetic phase transition, but rather indicate the formation of magnetic clusters, the spins of which are rapidly relaxing on the Mössbauer time scale. The splitting of the Mössbauer pattern for each magnetic site is characterized by the magnetic hyperfine field $B_{hf}$, which in the case of such nearly cubic materials reflects the ordered magnetic moment of the Fe sites. In agreement with the NP and susceptibility data, a single hyperfine pattern is discernible in the Mössbauer spectra below 140 K (AF1 phase), whereas below 67 K a second hyperfine pattern with increased $B_{hf}$ appears (AF2 phase). The relative intensity of the AF1 subspectrum decreases with decreasing temperature, but phase coexistence is still apparent down to 30 K. The 5 K spectrum can be described by a single, quite sharp six-line pattern, whereas the NP data show a residual AF1 fraction of 15 % even at 2 K. The fact that a separate sextet is no longer seen verifies that the Fe magnetic moment in the AF1 phase becomes equal to that of the AF2 phase at low *T*.

The isomer shifts (*IS*) of 0.45 mm/s at 295 K and 0.57 mm/s at 5 K are in the upper range of values typical for $Fe^{3+}$ [23], which suggests that the metal ions in $Sr_2FeOsO_6$ can essentially be described as $Fe^{3+}$ ($t_{2g}^3 e_g^2$) and $Os^{5+}$ ($t_{2g}^3$) sites, which is in agreement with the structural data. The presence of $Fe^{4+}$ in $Sr_2FeOsO_6$, as recently suggested by Feng et al. [21], can be discarded. The change in the low-*T* isomer shift values in the series $Sr_2FeB'O_6$ from 1.2 mm/s for B' = W [16] to 0.71 mm/s for B' = Re [17] to 0.57 mm/s for B' = Os reflects the successive stabilization of the $B'$ $t_{2g}$ with respect to the Fe $t_{2g}$ orbitals. Concomitantly, the iron oxidation state evolves from +2 to +3, where in the metallic ferromagnetic $Sr_2FeReO_6$ an intermediate-valence situation occurs [24], similar as in the prototypical $Sr_2FeMoO_6$ [25]. The large $B_{hf}$ of 49 T at 5 K confirms that the iron oxidation state in $Sr_2FeOsO_6$ is essentially +3. The phase transition from the AF1 to the AF2 phase below 67 K is only associated with a change in $B_{hf}$, whereas the *IS* values for the two phases are the same [22]. Thus, the size of the ordered magnetic moments changes, but the valence state of the Fe atoms remains unaffected. The Mössbauer signals of the AF2 phase are sharp, confirming a well-ordered magnetic state. In contrast, they are broader in the AF1 patterns and can be described by $B_{hf}$ distributions (see Fig. 2c). In particular, a broad background with a width of $\Delta B_{hf} \sim 10$ T suggests a high degree of spin disorder and/or spin fluctuations in the AF1 phase. The $T$ dependence of $B_{hf}$ follows the behavior of the total moment $\mu$ (cf. Figs. 2b and 1b) and confirms the gradual evolution of the Fe moments in the AF1 phase.

In order to understand these two spin structures of $Sr_2FeOsO_6$, we have investigated their electronic structure by performing spin-polarized DFT calculations [26] within the generalized gradient approximation [27]. First, we adopt the experimental lattice parameters measured by neutron diffraction at 2 K and optimize the atomic positions with the AF1 ordering that has higher symmetry ($I4/m$) than the AF2 ordering ($I4$). Based on this optimized atomic structure, we compare the total energies of AF1 and AF2 and find that AF2 is 10 meV lower in energy. Second, we re-optimize the above structure with the AF2 ordering and find that the total energy decreases by 40 meV more, so that the fully relaxed AF2 configuration is 50 meV more stable than AF1 in total. The resulting Fe-Os distances are $d_1 = 4.00$ Å and $d_2 = 3.94$ Å, which are well consistent with the experimental values. These results suggest that both, the magnetic exchange and lattice distortions favor the AF2 structure. The band structures, similar to $Sr_2CrOsO_6$ [28], show an energy gap ($E_g$) between the occupied Os-$t_{2g}$ ($t_{2g}^3$) and empty Fe-$t_{2g}$ and Os-$e_g$ states with $E_g = 0.10$ eV for AF1 and $E_g = 0.37$ eV for AF2 [22]. The gap is further enhanced by an additional $U$-term which accounts for electron correlation effects [20]

The variation in the Fe-Os distances along $c$ reflects a change of the orbital ordering (Fig. 3), which leads to a splitting of every $d$-band in AF1 into two bands in AF2. The structural distortion results in a modulation of the ratio $J_1/J_2$ of nearest neighbor Fe $3d$ – Os $5d$ and next nearest neighbor Os $5d$ – Os $5d$ interactions along $c$ (Fig. 1e). The shorter Fe-Os distances favor in every second iron-osmium double layer a spin arrangement with antiparallel alignment of Fe-Os and parallel alignment of Os—Os moments along $c$. This section of the spin structure is similar to the ferrimagnetic spin structure of $Sr_2CrOsO_6$ [13]. In the Fe-Os double layers with larger Fe-Os distances the spin arrangement of the AF1 phase with parallel Fe-Os and antiparallel Os-Os moments is retained. The combination of the two spin alignment schemes in AF2 reflects the competition between Fe-Os and Os-Os interactions. The spin structure adopted in such a situation sensitively depends on the ratio $J_1/J_2$ [15]. Both spin structures are prone to frustration, which is, however, relaxed by the structural distortion. This view is supported by the sharper Mössbauer spectra of AF2. In both spin structures the magnetic moments are aligned along the unique tetragonal axis. Another type of ordering with an antiferromagnetic alignment of ferromagnetic (111) planes was proposed for the insulating $Fe^{2+}/W^{6+}$ phase $Sr_2FeWO_6$ from model Hamiltonian calculations [29], but here only the $Fe^{2+}$ sublattice is magnetic.

In conclusion, a subtle interplay among competing exchange and electron-lattice interactions determines the complex magnetism of $Sr_2FeOsO_6$. The competing magnetic correlations occurring even above $T_N$ are reflected in the divergence between zero-field cooled (ZFC) and field cooled (FC) susceptibility data (Fig. 2d) as well as in the deviations from Curie-Weiss behavior [20,21]. In the paramagnetic regime, ferromagnetic interactions are dominant in spite of the overall antiferromagnetic order. The competing interactions explain why the magnetic order in the AF1 phase appears to be still incomplete, as suggested by the broad Mössbauer hyperfine patterns and the gradual evolution of magnetic moments. In this respect, $Sr_2FeOsO_6$ is comparable to the geometrically frustrated $Sr_2YRuO_6$ [30,31]. The increasing tetragonal distortion accompanying the formation of the AF1 phase (Fig. 1d) indicates that dynamic bond alternation may have already developed in AF1. This fine-tunes the exchange interactions and may finally trigger the transition to the AF2 magnetic structure below 67 K. The nature and strengths of the various interactions remain to be clarified in future work. Previous studies on the electronic structure of the $Sr_2CrB'O_6$ series [28,14] suggested that an interplay between the HD mechanism and superexchange interactions between local moments determines the magnetic properties [14]. The HD mechanism, which stabilizes the ferromagnetic metallic state in $Sr_2FeReO_6$, will be particularly important if the minority spin $t_{2g}$ orbitals of $B'$ and Fe or Cr have comparable energies. This mechanism is possibly less important in $Sr_2FeOsO_6$, as the Os $t_{2g}$ orbitals are lower in energy than the Re $t_{2g}$ orbitals, but it may still be enhanced compared to that in the Cr analogue $Sr_2CrOsO_6$. Because of the persistence of a nonmetallic state with local moments in the paramagnetic phase and the large $U$ term expected for high-spin $Fe^{3+}$ ($t_{2g}^3 e_g^2$), $Sr_2FeOsO_6$ may turn out to be a Mott-type insulator, as was suggested for $Sr_2CrOsO_6$ [15]. Our studies on $Sr_2FeOsO_6$ provide a critical test for models of the electronic behavior of double perovskites.

B.Y. and C.F. acknowledge financial support from the European Research Council Advanced Grant (ERC 291472).

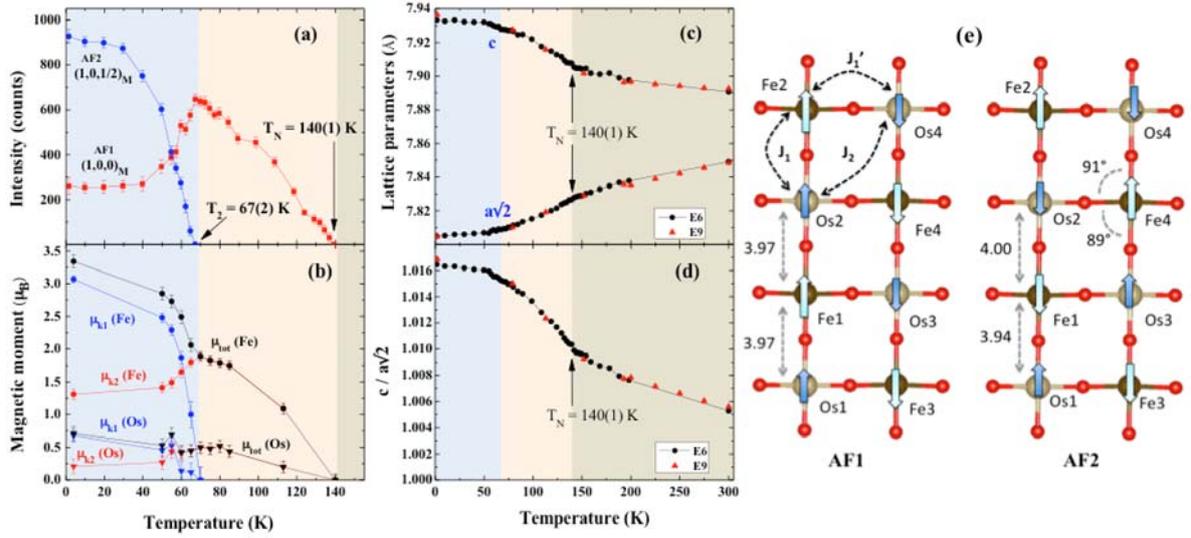

FIG. 1. (color online). Temperature dependence of (a) the intensity of the magnetic reflections $(1,0,0)_M$ and $(1,0,1/2)_M$ and (b) the magnetic moments of the Fe and Os atoms of the two magnetic phases AF1 and AF2 of $Sr_2FeOsO_6$. The filled black circles represent the total magnetic moments of both metal atoms. In (c) and (d), the temperature dependence of the lattice parameters as well as the $c/a$ ratios are shown in the cubic setting ($a_c \times b_c \times c_c = a_t\sqrt{2} \times a_t\sqrt{2} \times c_t$). In (e), the two magnetic structures are illustrated. Bond distances and angles obtained from DFT calculations as well as the dominant exchange pathways are indicated.

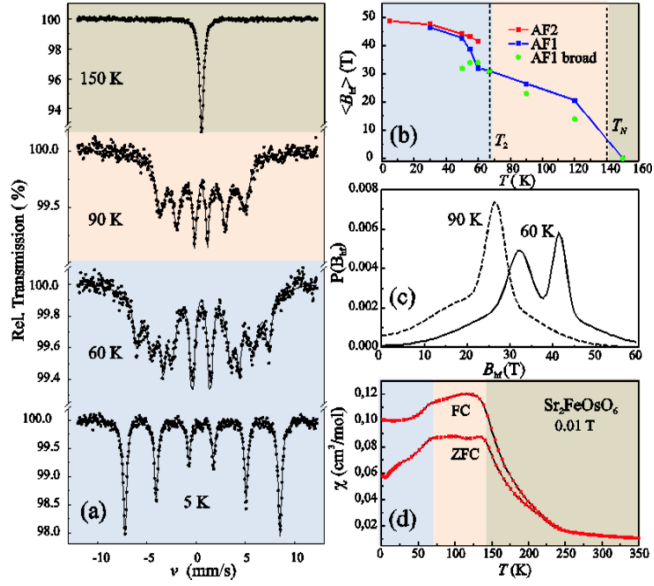

FIG. 2 (color online) (a) Mössbauer spectra of $Sr_2FeOsO_6$ at 5, 60, 90, and 150 K. (b) Temperature dependence of the average magnetic hyperfine fields $B_{hf}$. Solid lines are guides for the eye. (c) Hyperfine field distributions at 60 and 90 K obtained with a Voigt-based fitting algorithm. (d) Temperature dependence of the magnetic susceptibility in a field of 0.01 T.

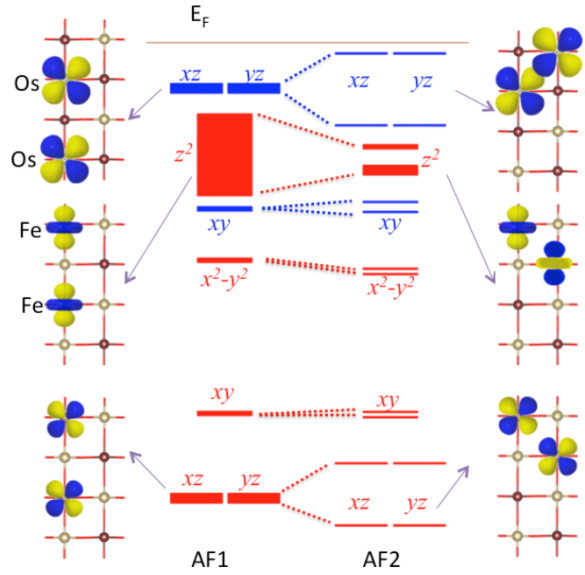

FIG. 3 (color online) The *d*-orbital diagram of the valence states of $Sr_2FeOsO_6$ in the AF1 and AF2 phases as calculated from the band structures [22]. The blue and red boxes represent Os states and Fe states, respectively. The positions of the boxes stand for the energies of the corresponding *d* bands, and the widths of the boxes indicate the bandwidths along the $k_z$ direction. Every *d*-band in AF1 splits into two bands in AF2, which correspond to the bonding and antibonding states. The wave functions of the Os $5d_{xz,yz}$, Fe $3d_z^2$, and Fe $3d_{xz,yz}$ states are illustrated as examples.